\documentclass[useAMS,usenatbib,usegraphicx]{mn2e}
\usepackage{color}

\newcommand{\kms}{km s$^{-1}$}

\title[Estimation of the SZ effects with unbiased multifilters]{The 
  estimation of the SZ effects
  with unbiased multifilters}
\author[Herranz et al.]{D. Herranz$^{1}$\thanks{E-mail:
    diego.herranz@isti.cnr.it}, J.L. Sanz$^{2}$, R.B. Barreiro$^{2}$ and
  M. L\'opez-Caniego$^{2}$\\
  $^{1}$Istituto di Scienze e Tecnologie dell'Informazione ``A. Faedo'', CNR, 
  via Moruzzi 1, 56124 Pisa, Italy \\
  $^{2}$Instituto de F\'\i sica de Cantabria, avda. Los Castros s/n, 
  39005 Santander, Spain}
\begin{document}

\date{Accepted --. Received --; in original form --}

\pagerange{\pageref{firstpage}--\pageref{lastpage}} \pubyear{2004}

\maketitle

\label{firstpage}

\begin{abstract}
In this work we study the performance of linear multifilters
for the estimation of the amplitudes of the thermal and kinematic
Sunyaev-Zel'dovich effects. 
We show that when both effects are
present, estimation of these effects with standard matched
multifilters is intrinsically biased. This bias is due to the fact
that both signals have basically the same spatial profile. We
find a new family of multifilters related to the matched multifilters
that cancel this systematic bias, hence we call them
Unbiased Matched Multifilters. We test the unbiased matched multifilters
and compare them with the standard matched multifilters using
simulations that reproduce the future Planck mission's observations.
We find that in the case of the standard matched multifilters 
the systematic bias in the estimation of the kinematic Sunyaev-Zel'dovich
effect can be very large, even greater than the statistical error bars.
Unbiased matched multifilters cancel effectively this kind of bias.
In concordance with other works in the literature,
our results indicate that the sensitivity and resolution of Planck
will not be enough to give reliable estimations of the kinematic
Sunyaev-Zel'dovich of individual clusters. 
However, since the estimation with the unbiased
matched multifilters is not intrinsically biased, it can be possible
to use them to statistically
study peculiar velocities in large scales using large sets of clusters.
\end{abstract}


\section{Introduction}

The Sunyaev-Zel'dovich effect~\citep{sz70}
is one of the most interesting and promising observational 
tools for cosmology.
During the last four decades, many works have addressed its 
usefulness as a probe that can be used to determine cosmological parameters
(in special when combined with other observational diagnostics 
such as gravitational
lensing or X-ray observations), to study the abundances and evolution of galaxy
clusters up to high redshifts, to estimate the gas mass fractions 
inside clusters,
to measure peculiar velocities of the gas and to map the inner structure of 
galaxy clusters.
Some excellent recent reviews on the physics of the SZ effect
can be found in \citet{reph95} and \citet{bir99}. 
Recent observational results have been
reviewed in~\citet{bir99} and~\citet{car00}. 
Another excellent review that focus in the application
of SZ effect observations to cosmology can be found in
\citet{car02}.

While the theory of the SZ effect physics is 
very well understood,
from the observational point of view its study
is still difficult.
This is
mainly due to 
the faintness of the effect --specially the elusive kinematic SZ effect-- 
and the presence of many
astrophysical ``contaminants'' in the frequencies where
the SZ effect is better observed.

The SZ effect appears as 
a secondary anisotropy in the Cosmic Microwave Background (CMB), that is, as a 
small spectral distortion in the CMB spectrum caused by the scattering of CMB
photons due to high energy electrons. Therefore, it must be disentangled from the
CMB fluctuations themselves, a task that is relatively easy to accomplish in
the case of the thermal SZ effect but very difficult in the case of the kinematic SZ.
Besides, all the contaminants (or ``foregrounds'') that affect CMB observations
affect as well SZ effect experiments: extragalactic point sources and Galactic
foregrounds such as dust, synchrotron and free-free emissions. 
As we will show later in this work, even the two different types of SZ effects,
the thermal and the kinematic, can be considered as contaminants one of the other.

In the last few years several SZ experiments have reached the 
sensitivities and angular resolutions needed to, for the first time in history,
fully exploit the power of SZ observations. Even more experiments are under construction
or planned for the next years \citep{car00,car02,nabila04a}.
Ever for these high-sensitivity experiments
the observation of the SZ effect is not an easy task,
and therefore a great care must be placed in
the analysis and interpretation of the data.

Several different approaches lead to the many different SZ data analysis
methods proposed in the literature. One approach is the \emph{component
separation}, which goal is to extract from the data all the different 
signals that were present for their separate study. In this context 
the SZ effect
is just one of the products of the separation.
Component separation techniques
typically make use of the different statistical distribution and 
spectral behaviour (i.e. the different frequency
dependence) of the components. An example of component
separation method that has been applied to simulations containing
SZ effect is the 
Maximum Entropy Method, e.g. \citet{hob99}. As noted by \citet{chem02} and
\citet{her02}, component separation methods can be very powerful
but there is a risk when the main goal is to study only one of the 
sources (in this case the SZ effect): an error in the separation of any one of
the components is easily propagated to the others, and therefore it is necessary
to be extremely careful that all the assumptions made (such as the frequency dependence of
the components, or their statistical independence, etc) are 
correct.

Another approach focuses only on the one component under study
(the SZ in this case), trying to extract it from the data or to
estimate a few parameters that characterise it. For example,
in the case of the SZ effect we can try to obtain a map of the Compton parameter
or just to obtain the positions of the clusters, their integrated fluxes, their
peculiar velocities, etc. We will call this approach \emph{detection} 
and/or \emph{estimation}. Detection/estimation techniques consider all the
other components of the data apart from the one under study as interference (noise)
and try to minimise its impact. 
Compared with component separation techniques, they tend to be
more robust in the sense that they do not need to model in detail
all the different components that form part of the ``background''.
Examples of this type of techniques applied
to SZ data are Bayesian 
non-parametric mapping of the Compton parameter \citep{chem02}, 
optimal 
filtering \citep{hael96,agh97,her02}, 
Markov Chain Monte Carlo (MCMC) sampling techniques \citep{hob03}
and parameter estimation via
simulated annealing \citep{han04a}.

The choice among the above mentioned techniques is a delicate issue; all of them
have some desirable properties and some drawbacks, and depending on the case
under study one or another (or even a combination of several of them) will be more
appropriate.
The good thing is that they are not mutually exclusive and in certain cases
they can work together to obtain better results. In general non-linear methods
such as MCMC and simulated annealing are more powerful, but they 
are not free of problems. MCMC methods are 
expensive and time-demanding from the computational point of view.
This make them unpractical for large blind SZ surveys.
The SASZ simulated annealing code \citep{han04a} is fast but in
presence of astrophysical contaminants such as extragalactic point
sources and Galactic foregrounds it is likely to suffer from
systematic errors that can be relatively large \citep{nabila04}.

Linear ``optimal'' filters may be not so powerful as non-linear 
detection/estimation techniques, but they have some nice properties 
that may make us consider their use. They are simple to understand and
implement. They are quite fast and not computationally demanding. Moreover,
since they do not need a detailed modelling of the background statistical
properties --apart from the fact that it is statistically homogeneous and that
it can be properly described by second-order statistics, namely the power 
spectrum-- they are very robust. Moreover, convolutive filters are appropriate to select 
``compact'' features (as clusters of galaxies) 
in the data by filtering out the large scales 
where diffuse components manifest and the small scales where pixel noise 
appear.
The multifilters (filters that operate
simultaneously on all the frequency channels of an experiment, taking
into account the frequency dependence of the SZ effect and the 
correlations among different channels in order to optimise the detection
of the SZ effect) presented in \citet{her02}
provide a powerful denoising tool (they minimise the background contamination)
as well as a straightforward amplitude estimation criterion.
In that work two families of multifilters, the matched multifilter (MMF) and
the scale-adaptive multifilter (SAMF) were studied and tested on realistic
Planck simulations.
The results showed that both families of multifilters lead to unbiased
estimators of the central value of the Compton parameter of the clusters 
in the simulations, even in presence of realistic contaminants. 
MMF give higher gain factors (the ratio between the variance of the 
background before and after filtering) than the SAMF, keeping the same level of
spurious detections, so they are preferred as our tool of choice.
The \citet{her02} multifilters were designed to detect/estimate the
amplitude of the thermal SZ effect. In the same way they can be constructed
to detect the more elusive kinematic SZ effect just by changing 
accordingly the frequency
dependence of the effect in the formulae that lead to the expression of the
filter.

In the previous discussion we have considered the case when 
only the thermal (or kinematic) SZ effect is present in the data. But if
the two effects appear at the same time, will be the multifilter estimator
still unbiased? The answer, as we will see, is no. Multifilters use information
on the frequency dependence and the spatial profile of the sources at the same time.
If two sources have different frequency dependence and profile they are likely to
be separated very well. But if the two sources have the same profile, or the same
frequency dependence, the problem will be somewhat degenerated and the estimator 
will show systematic effects.

A case of two sources with the same frequency dependence but different spatial
profile is evident: the kinematic SZ effect and the CMB spectra are identical
(except for second-order relativistic effects). This means that some residual
contamination from CMB fluctuations will remain after filtering, introducing errors
in the estimation of the kinematic SZ. Fortunately, the CMB fluctuations are damped
considerably on the angular scales of typical galaxy clusters (see for example
\citet{hu97}). Moreover, since CMB 
fluctuations can be symmetrically positive or negative,
on average the statistical systematic bias due to this effect is zero. 
Considered individually, however, all kinematic SZ 
measurements by mean
of multifilters will be affected by a non-removable error, 
and little can be done
to solve this problem.

On the other hand, the problem that appears when two signals with the same
spatial profile and different frequency dependence are overlapped can be solved. 
In our case these two signals are the thermal and the kinematic SZ effects. 
We will show here that their superposition leads inevitably to systematic errors in
the estimation of the SZ parameters when we use standard matched multifilters. This
errors can be very large in the case of the kinematic SZ effect and
can be very dangerous for automatic blind SZ surveys. This paper is devoted to
the study and cancellation of this dangerous type of bias.

The paper is structured as follows.
In section~\ref{sec:MMF}
we will briefly review the MMF and its properties. In section~\ref{sec:2signals}
we will show how the non-removable systematic bias appears in the MMF estimator when
two signals with the same spatial profile and location but different frequency dependence are
present. Section~\ref{sec:UMMF} shows how to design families of multifilters that
overcome that problem: the unbiased matched multifilters (UMMF). In section~\ref{sec:Planck}
we will test the MMF and the UMMF in realistic Planck simulations. Finally, we will
discuss our results in section~\ref{sec:discussion}.


\section{The standard Matched Multifilter} \label{sec:MMF}

The standard matched multifilter (MMF) was introduced by \citet{her02} and, independently,
by~\citet{nas02}.
Following~\citet{her02},
let us consider a set of $N$ astronomical images
corresponding to observations at different frequencies of a given region of the sky. A 
signal with a known frequency dependence is embedded in the data, so that the 
observations can be described with the following model:
\begin{equation} \label{eq:data_model}
d_{\nu}(\vec{x}) = f_{\nu} s_{\nu}(x) + n_{\nu}(\vec{x}), \ \ \ \nu=1,\ldots,N .
\end{equation}                                                                    
\noindent
where
the {\em generalised noise} $n_\nu(\vec{x})$ corresponds to the sum of the 
other emission components in the map. 
The background $n_{\nu}(\vec{x})$ is modelled as a homogeneous 
and isotropic random field with average value 
$\langle n_{\nu}(\vec{x})\rangle = 0$ and cross-power spectrum 
$P_{{\nu}_1 {\nu}_2}(q)$ ($q\equiv |\vec{q}|$) defined by
\begin{equation}  \label{eq:crosspower_def}
\langle n_{\nu_1}(\vec{q})n^*_{\nu_2}(\vec{q'})\rangle = 
P_{\nu_1\nu_2}(q) \, \delta_D^2 (\vec{q} - \vec{q'}),
\end{equation}
\noindent
where $n_{\nu}(\vec{q})$ is the Fourier transform of 
$n_{\nu}(\vec{x})$ and $\delta_D^2$ is the 2-D Dirac distribution.

The spatial profile $s_{\nu} (x)$ is usually written as the product of
the source amplitude $A$ and a spatial template $\tau (x)$.
In general, the template at each frequency $\nu_i$ is the result of the
convolution of the normalised profile of the source and the antenna beam
at that frequency.
Hereinafter, we assume for simplicity that the source has spherical symmetry, 
$x\equiv |\vec{x}|$, but the methods can be generalised for non-symmetric profiles. 
For simplicity as well we will assume that the source is centred at the position
$x=0$.

Equation (\ref{eq:data_model}) can be expressed in a more compact way as a 
vector equation,
\begin{equation} \label{eq:data_model_vectorial}
\mathbf{d}\mathnormal{(\vec{x})} = \mathnormal{A} \mathbf{F} \mathnormal{(x)} 
+ \mathbf{n}\mathnormal{(\vec{x})}.
\end{equation}
\noindent
where the elements of vector $\mathbf{F}$ are $F_{\nu} = f_{\nu} \tau_{\nu}$. 
Hereinafter, we will employ the usual boldface notation for the $N$-component
vectors in frequency whereas the ``arrow'' notation $\vec{x}, \ \vec{q} \ $ will refer
to 2-dimensional vectors in real space or Fourier space.

The goal is to obtain a set of $N$ linear filters that, once applied to the
data, allow us to determine the amplitude $A$ of the signal with a minimum
interference of the contaminant noise. 
The filters should take advantage on the knowledge of the frequency dependence
of the signal $\mathbf{f}$,
the characteristic profile of the source $\tau$ and the 
cross-power spectrum of the noise, that can be directly estimated from the data.
The filters are applied to the $N$ frequency
channels in order to give a quantity --called \emph{filtered map}-- at each 
position that will be used as an estimator of $A$. Let $\Upsilon (\vec{x}) = \{ \upsilon_{\nu_1}(\vec{x}),
\ldots,\upsilon_{\nu_N}(\vec{x}) \}$ the vector of $N$ filters we are looking for, then the 
filtered map at the position $\vec{b}$ is given by
\begin{equation} \label{eq:filtered_map_space}
w(\vec{b}) = \sum_{\nu} \int d \vec{x} \  d_{\nu} (\vec{x}) \upsilon_{\nu} \left( \left| \vec{x}-
\vec{b} \right| \right),
\end{equation}
\noindent
that can be expressed in Fourier space in a very compact way
\begin{equation} \label{eq:filtered_map_Fourier}
w(\vec{b}) = \int d \vec{q} \  e^{-i \vec{q} \vec{b}}  
\mathbf{d}^{\mathrm{t}}(\vec{q}) \Upsilon (\mathnormal{q}),
\end{equation}
\noindent
where the superscript $t$ denotes vector transposition, and the operation inside the integral
is the usual scalar product of vectors. The variance of the filtered map
can be expressed as
\begin{equation} \label{eq:variance}
\sigma^2_w = \langle w^2(\vec{b}) \rangle  - \langle w(\vec{b}) \rangle^2 =
\int d \vec{q} \ \Upsilon^{\mathrm{t}} \mathbf{P} \Upsilon.
\end{equation}
\noindent
The explicit dependence in $q$ has been removed from the previous equation for 
the sake of simplicity of notation. Unless it is necessary to include it, we will
do the same in the following.
 
The MMF is obtained when two constraints are imposed on the filters $\Upsilon$:
\begin{enumerate}
\item That $w(\vec{0})$ is an \emph{unbiased estimator} of the amplitude of the source,
that is,  $w(\vec{0})=A$.
\item That $w(\vec{0})$ is an \emph{efficient estimator} of the amplitude of the source,
that is, the variance of the $w$ values is minimum.
\end{enumerate}

As was shown in \citet{her02}, the filters that satisfy these conditions are given
by:
\begin{equation} \label{eq:MMF}
\Upsilon_{MMF} = \alpha^{-1} \mathbf{P}^{\mathnormal{-1}} \mathbf{F}, \ \ \ \mathnormal{\alpha} = 
\int d \vec{q} \ \mathbf{F}^{\mathrm{t}} \mathbf{P}^{\mathrm{-1}}  \mathbf{F},
\end{equation}
\noindent
where $\mathbf{P}^{\mathrm{-1}}$ is the inverse matrix of the cross-spectrum
$\mathbf{P}$. Using eq. (\ref{eq:variance}) we have that for the MMF
\begin{equation}   \label{eq:variance_MF}
\sigma^2_w = 
\int d \vec{q} \ \Upsilon_{MMF}^{\mathrm{t}} \mathbf{P} \Upsilon_{MMF} = \alpha^{-1}.
\end{equation}

\section{Signals with the same spatial profile and different frequency dependence} \label{sec:2signals}

The MMF is designed to produce an unbiased and efficient linear estimator of the amplitude of 
a signal with known frequency dependence and spatial profile 
in  multiwavelenght observations. But what
happens when there are present two signals with the same spatial profile and location but different
frequency dependence? Such is the case, for example, of the thermal and kinetic Sunyaev-Zel'dovich
effects. Both effects are superimposed with identical spatial distribution (if we do not
take into account intracluster gas motions that can distort the shape of the kinematic effect)
but quite different frequency dependence.

\subsection{Thermal and kinematic Sunyaev-Zel'dovich effects} \label{sec:SZ_effects}

The thermal SZ effect is due to the inverse Compton scattering 
of CMB photons by free electrons in the hot intra-cluster gas. It produces a shift
in the CMB spectrum to higher frequencies in the direction where the cluster
is observed. This effect is characterised by the comptonization parameter $y_c$,
\begin{equation} \label{eq:compton} 
y_c = \frac{k_B \sigma_T}{m_e c^2} \int T_e(l) n_e(l) dl ,
\end{equation}
\noindent
where $k_B$ is the Boltzmann constant, $\sigma_T$ the Thomson section, $m_e$ the electron
mass, $c$ the velocity of light, $T_e$  and $n_e$ are the  temperature and density 
of the electrons in the gas and $l$ is the distance along the line of sight. Ignoring
relativistic effects, the frequency dependence of the thermal SZ effect is given by
\begin{equation} \label{eq:SZt_freqdep}
\left( \frac{\Delta T}{T} \right)_{SZt} (\nu)  = y_c \ f(\tilde{\nu}), \ \ \ 
\tilde{\nu} = \frac{h \nu}{k_B T_{CMB}}, 
\end{equation}
\noindent 
where $h$ is the Planck constant and $T_{CMB}$ the temperature of the CMB, and where
\begin{equation} \label{eq:SZt_freqdep2}
f(x) = x \frac{e^x+1}{e^x-1} - 4.
\end{equation}

The kinematic SZ effect, on the other hand, is due to the Doppler effect
that arises because of the radial peculiar velocity of the cluster $v_r$ along the
line of sight. The intensity of the kinematic effect, not taking into account
second-order relativistic corrections and considering $T_e$ to be constant
along the cluster, is
\begin{equation} \label{eq:SZk_freqdep}
\left( \frac{\Delta T}{T} \right)_{SZk} (\nu)  = -  y_c \
\frac{v_r m_e c}{k_B T_e},
\end{equation}
\noindent
and the frequency dependence is (again ignoring relativistic corrections)
constant in $\Delta T/T$ units. The intensity of the kinematic effect
is in general significantly smaller than the intensity of the thermal effect.

At high frequencies, the expressions given above are not accurate and 
relativistic corrections should be taken into account when dealing with real data.
In this work we restrict the discussion to the non-relativistic case for simplicity.

\subsection{Bias in the MMF estimator}

Combining the two SZ effects our data model is 
\begin{equation} \label{eq:data_model_both_vectorial}
\mathbf{d} \mathnormal{(\vec{x})} = 
\left( \frac{\Delta T}{T} \right)_{SZt+k} =
y_c \left[ \mathbf{F}  \mathnormal{(\vec{x})} -
 V \mathbf{\tau} \mathnormal{(\vec{x})}  \right]
+\mathbf{n} \mathnormal{(\vec{x})} ,
\end{equation}
\noindent
where $V = (v_r m_e c) / (k_B T_e)$  
and the vector $\mathbf{F}$ is
constructed using the profile $\tau$ and the thermal SZ frequency dependence
in eqs. (\ref{eq:SZt_freqdep}) and (\ref{eq:SZt_freqdep2}).

Imagine a MMF designed for the detection of the thermal SZ
effect is applied to the data in (\ref{eq:data_model_both_vectorial}).
The filter will be the one given in eq. (\ref{eq:MMF}) using the vector $\mathbf{F}$
that corresponds to eq. (\ref{eq:data_model_both_vectorial}).
It is easy to calculate the average value of the 
filtered map:
\begin{eqnarray} \label{eq:filtered_map_MMFt}
\langle w_t(\vec{0}) \rangle = \int d \vec{q} \ 
y_c \left[  \mathbf{F}^{\mathrm{t}} -V  \mathbf{\tau }^{\mathrm{t}}  \right] 
\Upsilon_{MMFt} = \nonumber \\
y_c \ \alpha^{-1} \int d \vec{q} \ \mathbf{F}^{\mathrm{t}} \mathbf{P}^{\mathnormal{-1}} \mathbf{F} -
y_c  V \alpha^{-1} \int d \vec{q} \ \mathbf{\tau}^{\mathrm{t}} \mathbf{P}^{\mathnormal{-1}} \mathbf{F}
 = \nonumber \\
y_c - y_c V \frac{\beta}{\alpha},
\end{eqnarray}
\noindent
where the constant $\beta$ is defined as
\begin{equation} \label{eq:beta}
\beta = \int d \vec{q} \ \mathbf{\tau}^{\mathrm{t}} \mathbf{P}^{\mathnormal{-1}} \mathbf{F} .
\end{equation}

According with eq. (\ref{eq:filtered_map_MMFt}), due to the presence of the kinematic SZ
effect the estimator $w_t(\vec{0})$ is not longer unbiased, and the estimation of $y_c$ 
will have a systematic error proportional to $V \beta / \alpha$.

The same will occur when a MMF designed for the detection of the kinematic SZ effect.
In  that case, the frequency dependence to be used is a constant whose
value can be included in $V$ and therefore $f_{\nu} = 1$. The shape of the MMF is then
\begin{equation} \label{eq:MMFk}
\Upsilon_{MMFk} = \gamma^{-1} \mathbf{P}^{\mathnormal{-1}} \mathbf{\tau}, \ \ \ \mathnormal{\gamma} = 
\int d \vec{q} \ \mathbf{\tau}^{\mathrm{t}} \mathbf{P}^{\mathrm{-1}}  \mathbf{\tau},
\end{equation}
The average value of the 
filtered map is then
\begin{eqnarray} \label{eq:filtered_map_MMFk}
\langle w_k(\vec{0}) \rangle = \int d \vec{q} \ 
y_c \left[  \mathbf{F}^{\mathrm{t}} -V  \mathbf{\tau }^{\mathrm{t}}  \right] 
\Upsilon_{MMFk} = \nonumber \\
y_c \ \gamma^{-1} \int d \vec{q} \ \mathbf{F}^{\mathrm{t}} \mathbf{P}^{\mathnormal{-1}} \mathbf{\tau} -
y_c  V \gamma^{-1} \int d \vec{q} \ \mathbf{\tau}^{\mathrm{t}} \mathbf{P}^{\mathnormal{-1}} \mathbf{\tau}
 = \nonumber \\
y_c \frac{\beta}{\gamma}  - y_c V.
\end{eqnarray}
Note that since $y_c$ and $V$ appear as a product in eq. (\ref{eq:data_model_both_vectorial}),
$\langle w_k(\vec{0}) \rangle$ is not a direct estimator of $V$ but of the quantity $y_c V$.
Dividing by $y_c$ and changing the signs we obtain
\begin{equation} \label{eq:bias_MMFk}
 - \frac{\langle w_k(\vec{0}) \rangle }{y_c} = V - \frac{\beta}{\gamma} , 
\end{equation}
\noindent
and therefore we find that our estimator is again biased.

As we will see later in section~\ref{sec:Planck}, in the case of microwave observations 
of the SZ effect, the bias in the MMF estimator of the thermal SZ effect is normally
negligible, but in the case of the kinematic SZ effect it is expected to be very
large, of the order of the effect itself or even bigger. Therefore, this bias must be
taken into account and corrected.

\subsection{Cancelling the bias}  \label{sec:cancelling}

There are several ways in which the bias described above can be taken into account.
The first option is simply to separate the thermal and kinematic effects before estimating
their amplitudes. This goal can be attained using a convenient component separation technique,
for example the Maximum Entropy Method \citep{hob99},
non-parametric Bayesian detection \citep{chem02} or MCMC sampling techniques \citep{hob03}.
The problem is that in practise
no component separation technique works perfectly, and some uncontrolled residuals
always remain that will still bias the estimation of the amplitude of the effects.

A second option consists in using eqs. (\ref{eq:filtered_map_MMFt}) and 
(\ref{eq:bias_MMFk}) to subtract the bias factor in each case. This is a perfectly 
licit option 
in principle, but can be a little difficult to carry out in a real 
case. The argument is as follows: to determine the bias in $y_c$ we need, according to
eq. (\ref{eq:filtered_map_MMFt}), to know the value of $y_c V$, whose estimation is biased. 
To remove the bias in $y_c V$ we need, according to eq. (\ref{eq:bias_MMFk}), to know the value
of $y_c$, which is biased as well. 
Moreover, on an individual basis (cluster by cluster) the quantities $V$ and $y_c$ will
be estimated with a non-zero error.
Therefore, trying to cancel the bias in this way leads us to
a vicious circle. In section~\ref{sec:Planck} we will see that for the case of the  future
Planck satellite observations the bias in $y_c$ will be small, and the vicious circle
could be circumvented just by considering the estimated $y_c$ as the correct one. This 
would however introduce further errors in any case. Besides, in other experiments
the bias in $y_c$ can be not so small as in the Planck case is. Therefore, we do not
recommend this option to cancel the bias of the MMF. In the next section we will
describe two novel lineal multifilters that automatically cancel the bias
in the estimation of the thermal and kinematic SZ amplitudes.

\section{Unbiased Matched Multifilter} \label{sec:UMMF}

Instead of following any of the approaches described in section~\ref{sec:cancelling},
let us see if it is possible to design linear filters that are nearly as efficient as
the MMF and do not show the systematic bias described above. To do it we will make
use of the knowledge on the two distinct frequency dependences of the thermal and
kinematic SZ effects. The resulting filters will be called \emph{unbiased matched
multifilters} (UMMF). Specifically, we search for a set of filters specifically
designed to give the thermal amplitude $y_c$ without bias and a different 
set of filters  
 specifically
designed to give the kinematic amplitude $y_c V$.

\subsection{Unbiased Matched Filter for the thermal SZ effect}  \label{sec:UMMFt}

We construct the filters $\mathbf{\Psi} \mathnormal{=} \{ \psi_{\nu_1}, \ldots ,  \psi_{\nu_N} \}$
so that the filtered map $w_{\Psi}$ satisfies the canonical matched filter
conditions plus another one aimed at cancelling the bias due to the presence of 
the kinematic effect. 
\begin{enumerate}
\item The filters, when applied to the thermal part of eq. (\ref{eq:data_model_both_vectorial}),
  give, at the position of the source, the value of the comptonization parameter, that is,
  \begin{equation}  \label{eq:condition1_UMMFt}
    \int d \vec{q} \ \mathbf{F}^{\mathnormal{t}} \mathbf{\Psi} = 1 ,
  \end{equation}
\item The filters, when applied to the kinematic part of eq. (\ref{eq:data_model_both_vectorial}),
  give, at the position of the source, no contribution to the filtered map, that is,
  \begin{equation}  \label{eq:condition2_UMMFt}
    \int d \vec{q} \ \mathbf{\tau}^{\mathnormal{t}} \mathbf{\Psi} = 0 ,
  \end{equation}
\item The variance of the filtered map, $\sigma_{w_{\Psi}}$, is minimum (\emph{efficient} estimator).
\end{enumerate}

The two first two conditions ensure that the filtered map at the position of the source 
is an unbiased estimator of the thermal SZ effect. The third one is the condition for efficiency
of the estimator. The filters that satisfy the previous three conditions are
\begin{equation} \label{eq:UMMFt}
\mathbf{\Psi} = \frac{1}{\Delta} \mathbf{P}^{\mathnormal{-1}} \left( \gamma \mathbf{F} - 
\beta \mathbf{\tau} \right),
\ \ \ \Delta=\alpha \gamma - \beta^2,
\end{equation}
\noindent
where the constants $\alpha$, $\beta$ and $\gamma$ are defined in equations (\ref{eq:MMF}),
(\ref{eq:beta}) and (\ref{eq:MMFk}). It is straightforward to see that 
$\langle w_{\Psi}(\vec{0}) \rangle = y_c$ and that the variance of the filtered map is
\begin{equation} \label{eq:variance_UMMFt}
  \sigma^2_{w_{\Psi}} = \frac{\gamma}{\Delta}.
\end{equation} 

\subsection{Unbiased Matched Filter for the kinematic SZ effect}  \label{sec:UMMFk}

Let us now find the filters $\mathbf{\Phi} \mathnormal{=} \{ \phi_{\nu_1}, \ldots ,  \phi_{\nu_N} \}$
so that the filtered map $w_{\Phi}$ satisfies the canonical matched filter
conditions plus another one aimed at cancelling the bias due to the presence of 
the thermal effect. 
\begin{enumerate}
\item The filters, when applied to the thermal part of eq. (\ref{eq:data_model_both_vectorial}),
  give, at the position of the source, no contribution to the filtered map, that is,
  \begin{equation}  \label{eq:condition1_UMMFk}
    \int d \vec{q} \ \mathbf{F}^{\mathnormal{t}} \mathbf{\Phi} = 0 ,
  \end{equation}
\item The filters, when applied to the kinematic part of eq. (\ref{eq:data_model_both_vectorial}),
  give, at the position of the source, the value of the product $y_c V$, that is,
  \begin{equation}  \label{eq:condition2_UMMFk}
    \int d \vec{q} \ \mathbf{\tau}^{\mathnormal{t}} \mathbf{\Phi} = 1 ,
  \end{equation}
\item The variance of the filtered map, $\sigma_{w_{\Phi}}$, is minimum (\emph{efficient} estimator).
\end{enumerate}

The two first two conditions ensure that the filtered map at the position of the source 
is an unbiased estimator of the kinematic SZ effect. The third one is the condition for efficiency
of the estimator. The filters that satisfy the previous three conditions are
\begin{equation}  \label{eq:UMMFk}
\mathbf{\Phi} = \frac{1}{\Delta} \mathbf{P}^{\mathnormal{-1}} \left( -\beta \mathbf{F} + 
\alpha \mathbf{\tau} \right),
\end{equation}
\noindent
where the constants $\alpha$, $\beta$, $\gamma$ and $\Delta$ 
are the same as in eq. (\ref{eq:UMMFt}).
It is straightforward to see that 
$\langle w_{\Phi}(\vec{0}) \rangle = y_c V $ and that the variance of the filtered map is
\begin{equation} \label{eq:variance_UMMFk}
  \sigma^2_{w_{\Phi}} = \frac{\alpha}{\Delta}.
\end{equation}

\section{Application to simulated Planck data} \label{sec:Planck}

As a test of the previous ideas, we now apply both the standard and the unbiased matched
multifilters to simulated Planck observations in order to estimate the thermal
and kinematic SZ of test clusters placed on the simulations. Our goal is just to
show by an example how the biases described in section~\ref{sec:2signals} appear and 
how the unbiased matched multifilters are able to cancel it. 
Therefore, to keep the example simple and clear we will
restrict ourselves to ideal conditions in which the spatial profile of the
clusters is perfectly known from the beginning. 
A full study of the performance of the filters in the Planck case,
including uncertainties in the cluster profile, asymmetric profiles
and realistic cluster distributions is out of the scope of this work
and will be addressed in the future.

We will focus specially on filters designed to extract the kinematic SZ effect.
Note that due to the extreme faintness of the kinematic SZ effect the task
of detecting it is going to be very difficult. In fact, Planck sensitivities, angular
resolution
and noise levels do not make it the best experiment to study the kinematic
SZ effect. In spite of this, we choose the Planck mission as our example 
scenario for two reasons. The first one is that  Planck's instrumental specifications,
noise levels and performance have been thoroughly studied in the literature,
and realistic simulations of Planck observations are relatively easy to 
produce and analyse. The second reason is that at Planck's angular resolution
most of the clusters will appear as point sources in the sky. Therefore,
their observed profile will be practically equal to the 
well-known beam profile of the Planck detectors.
Then, the assumption that the spatial profile of the clusters is known 
is not so unreasonable.

\subsection{Simulations} \label{sec:simulations}

Even though for this work we are going to use very simplistic toy clusters
and non-relativistic SZ effects, we intend to simulate the other
astrophysical and instrumental signals that constitute the
generalised noise $\mathbf{n} \mathnormal{(\vec{x})}$ in eq. 
(\ref{eq:data_model_vectorial})
in the most realistic way possible. 
Our simulations include the latest available information about 
the physical
components of the emission (CMB, Galactic foregrounds and extragalactic point sources)
and the technical specifications of the different Planck channels.
The simulations were performed in patches
of the sky of 
$12.8^{\circ} \times 12.8^{\circ}$. Table~\ref{tb:tech_mf} shows the assumed observational
characteristics of the simulated maps.
\begin{table} 
  \caption{Technical characteristics of the 9 simulated Planck channels.
    Column two lists the FWHM assuming a Gaussian beam. Column three shows
    the pixel size in arcmin adopted in our simulations.
    In column four the instrumental noise variance per pixel is given
    in $\Delta T/T$ units (thermodynamic temperature).}
  \label{tb:tech_mf}
  \begin{tabular}{c c c c}
    \hline
    Frequency & FWHM     & Pixel size & $\sigma_{noise}$ \\
    (GHz)     & (arcmin) & (arcmin)   & ($\Delta T/T$)   \\
    \hline
    30	& 33.0 	& 6.0   & 1.1$\times 10^{-5}$ 	\\
    44	& 24.0  & 6.0   & 1.1$\times 10^{-5}$   \\
    70	& 13.0  & 3.0   & 2.2$\times 10^{-5}$   \\
    100 & 9.2   & 3.0   & 6.1$\times 10^{-6}$   \\
    143	& 7.1   & 1.5	& 1.0$\times 10^{-5}$  	\\
    217	& 5.5	& 1.5	& 1.6$\times 10^{-5}$	\\
    353	& 5.0	& 1.5	& 4.9$\times 10^{-5}$	\\
    545	& 5.0	& 1.5	& 4.9$\times 10^{-4}$   \\
    857	& 5.0	& 1.5	& 2.2$\times 10^{-2}$   \\
    \hline
  \end{tabular}
\end{table}

The $C_{\ell}$'s for the CMB simulation were generated using the
CMBFAST code \citep{sel96}
for a spatially-flat $\Lambda$CDM Universe 
with $\Omega_m=0.3$ and $\Omega_{\Lambda}=0.7$
(Gaussian realisation).
The simulated Galactic emission includes four components:
thermal dust, spinning dust, free-free and synchrotron. 
The details of the Galactic foreground simulations are identical to
those described in \citet{her02} (see references therein). 
The extragalactic point source simulations come from
the model of \citet{T98} for a Gaussian realisation with the
same cosmological parameters used for the CMB simulation.

Synthetic clusters were simulated using the spatial
profile
\begin{equation} \label{eq:cluster_profile}
\tau (x) = \frac{r_c r_v}{r_c+r_v}
\left( \frac{1}{\sqrt{r_c^2+x^2}} - 
\frac{1}{\sqrt{r_v^2+x^2}} \right) .
\end{equation}
\noindent
The previous profile is a modification of the basic multiquadric
profile in which $r_c$ takes the role of the core radius of the
cluster and $r_v$ is a limiting cut scale that can be associated
with the virial radius of the clusters. For $x << r_v$ the profile
in eq. (\ref{eq:cluster_profile}) behaves like the classical $\beta$ model
(with $\beta=1/2$)
whereas for $x>>r_v$ the profile quickly drops to zero. The
profile in eq. (\ref{eq:cluster_profile}) is continuous,
gives a good approximation for the typical cluster profile and
is well-behaved in Fourier space.
For this work we used $r_v = 10 r_c$.

\subsection{Preliminary analysis of the simulations} \label{sec:preliminar}

Given a set of maps corresponding to the simulations
described above for the nine Planck frequencies, it is easy
to estimate the values of the cross-power matrix $\mathbf{P}$ and
therefore to calculate the values of integrals $\alpha$, $\beta$
and $\gamma$ for this case. Doing so, we obtain
the values
\begin{eqnarray} \label{eq:cte_values}
\alpha & = & 2.3721   \times 10^{10} \nonumber \\
\beta  & = & -1.4482  \times 10^{9} \nonumber \\
\gamma & = & 1.3694   \times 10^{9} \nonumber \\
\Delta & = & 3.0386   \times 10^{19} .
\end{eqnarray}
\noindent
According to these numbers and eqs. (\ref{eq:variance_UMMFt}) and 
(\ref{eq:variance_UMMFk}) we should expect that after using
the filters $\Psi$ and $\Phi$ 
the filtered maps should have variances 
$\sigma_{\Psi} \simeq 6.7 \times 10^{-6}$ and
$\sigma_{\Phi} \simeq 2.8 \times 10^{-5}$, in $\Delta T/T$ units.

Let us remember that for the case of the classical
matched multifilters the expected variances of the filtered
maps should be $\sigma_{MMFt} = (1/\alpha)^{1/2} \simeq 6.5 \times 10^{-6}$
and $\sigma_{MMFk} = (1/\gamma)^{1/2} \simeq 2.7 \times 10^{-5}$,
respectively. Then the addition of the bias cancellation constraint
in the formulation of the UMMF leads to a small increment of the
variance of the filtered maps, but a very small one indeed.

On the other hand, from equations 
(\ref{eq:filtered_map_MMFt})
and (\ref{eq:bias_MMFk}) it is possible to calculate
the expected biases that will arise with the standard MMF. Let us 
denote the estimation of a quantity $\xi$ with the standard MMF as
$\hat{\xi}$. We have for the case of the Planck simulations
that
\begin{eqnarray}  \label{eq:expected_biases}
\frac{\hat{y}_c}{y_c} & = & 1 + 0.06 V  ,  \nonumber \\
 \frac{\hat{V}}{V}    & = & 1 + \frac{1.05}{V} .
\end{eqnarray} 
\noindent
Depending on the value of $V$, some of the previous quantities 
can be considerable. Typical values are $V \leq 0.1 $, and
therefore we find that for the case of the thermal effect
the bias 
will be negligible ($\hat{y}_c/y_c \simeq 1$) whereas
for the kinematic effect the bias is huge ($\hat{V}/V \simeq 10$)
and can not be discarded.
\emph{Standard MMF for the estimation of the kinematic SZ effect 
is strongly biased}.

The expected errors in the estimation of the amplitudes of
the
thermal and kinematic effects by means of the multifilters
is directly related to the variances of the filtered map
calculated above. Note that in the case of Planck the 
variances in the maps filtered with ``kinematic-type'' multifilters
(both $\Phi$ filters and MMFk) are higher than the variances
for the ``thermal-type'' multifilters. That is bad news. It implies
that the detection and estimation of the kinematic effect by
means of multifilters is difficult not only due to the 
faintness of the effect, but also to the relatively high 
intrinsic variance of the filtered map.

\subsection{Test}

In order to test whether the numbers obtained above are
correct or not we ran a set of simulations, including synthetic
clusters with the profile (\ref{eq:cluster_profile}) and
the frequency law (\ref{eq:data_model_both_vectorial}).
We simulated 5000 of these clusters. In order to avoid cluster
overlapping only five cluster per simulated sky map were included.
For this test, we simulated very bright and big clusters,
with fixed parameters $r_c = 1.5$ arcmin, $y_c = 10^{-4}$ and 
$V = -0.1$. For a temperature of the electrons $T_e \simeq 5$ keV
that value of $V$ corresponds to a velocity along the line of 
sight of $v_r \simeq 300 \ \mathrm{km s^{-1}}$.
The simulations were filtered with standard matched multifilters
designed for the detection/estimation of both thermal and kinematic
SZ effects as well as with the $\Psi$ and $\Phi$ unbiased matched
multifilters.

We will focus here in the performance of the
filters regarding the \emph{estimation} of the SZ effects.
We will assume that the presence of the clusters is already
stablished and their locations well known. 
We would like to
remark that in a more general situation,
where detection must be achieved before estimating the
parameters, other sources of systematic bias different from the
one discussed in this work may arise due to the detection
criterion itself. For example, if detection is done by looking
for local maxima in the images, clusters located in areas
where the background is positive will be favoured for detection,
and then in the background will not have zero mean \emph{in
the areas where clusters are detected}, leading to a new
source of bias. For a more complete discussion on this, see for example
\citet{her02b} and \citet{can04}. These bias depend on the choice of
the detection device and 
their study
is out of the scope of this work.

\subsubsection{Performance of MMFt}

\begin{figure}
\includegraphics[width=84mm]{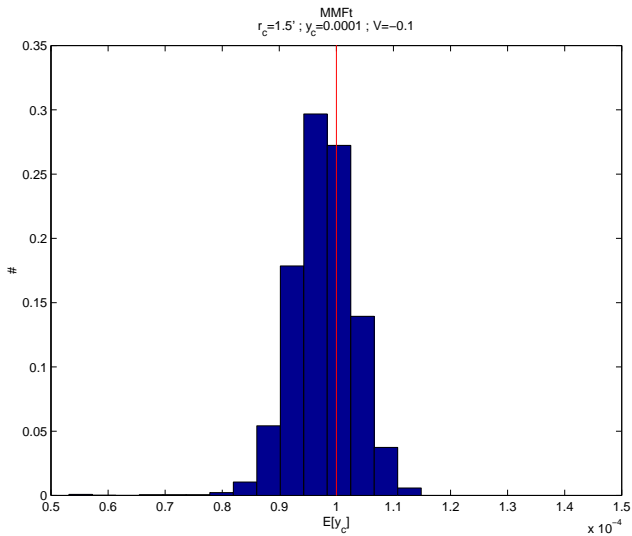}
\caption{Normalised histogram of the values of the estimated 
$y_c$ parameter using standard MMF for the thermal SZ effect.
The true value of $y_c$ is shown with a vertical red line.}
\label{fig:fig1}
\end{figure}

Figure~\ref{fig:fig1} shows
the results of the estimation of $y_c$ for the 5000
simulated clusters using the standard MMF for the thermal SZ effect.
The average value of the estimated $y_c$ is $\langle \hat{y}_c \rangle
= 9.77 \times 10^{-5}$ and the dispersion around this value is
$\sigma_{y_c} = 5.4  \times 10^{-6}$. The average standard deviation
of the filtered maps is, for comparison, $\sigma_{w_{MMFt}}=6.2 
\times 10^{-6}$. The value for this quantity 
that was predicted in section~\ref{sec:preliminar} was $6.5 \times 10^{-6}$.
Eq. (\ref{eq:expected_biases}) predicts $\hat{y}_c/y_c = 0.994$, and the 
value that is found in the simulations is $\hat{y}_c/y_c = 0.977$.

\subsubsection{Performance of $\Psi$ multifilters}

\begin{figure}
\includegraphics[width=84mm]{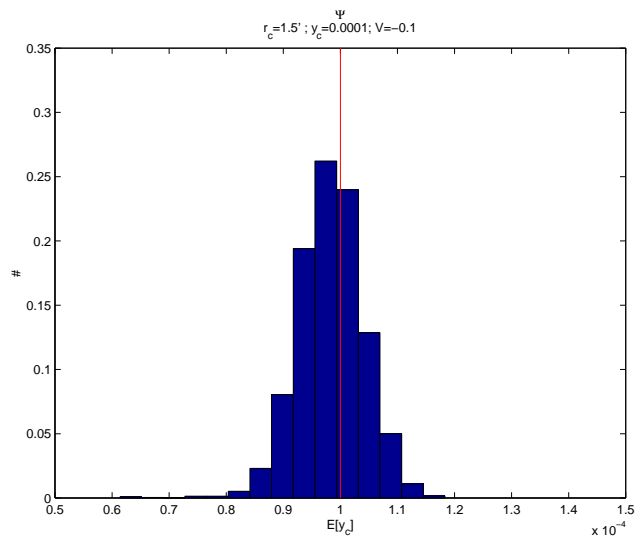}
\caption{Normalised histogram of the values of the estimated 
$y_c$ parameter using $\Psi$ UMMF for the thermal SZ effect.
The true value of $y_c$ is shown with a vertical red line.}
\label{fig:fig2}
\end{figure}

Figure~\ref{fig:fig2} shows
the results of the estimation of $y_c$ for the 5000
simulated clusters using the unbiased $\Psi$ multifilters
for the thermal SZ effect.
The average value of the estimated $y_c$ is $\langle \hat{y}_c \rangle
= 9.83 \times 10^{-5}$ and the dispersion around this value is
$\sigma_{y_c} = 5.7  \times 10^{-6}$. The average standard deviation
of the filtered maps is, for comparison, $\sigma_{w_{\Psi}}=6.5 
\times 10^{-6}$. The value for this quantity 
that was predicted in section~\ref{sec:preliminar} was $6.7 \times 10^{-6}$.
A small bias is present as well in this case, that can be
attributed to the finite size of the sample. Hence, for this case
both the $\Psi$ and the standard MMFt multifilters perform 
similarly well, down to the precision of the simulations.

\subsubsection{Performance of MMFk}

\begin{figure}
\includegraphics[width=84mm]{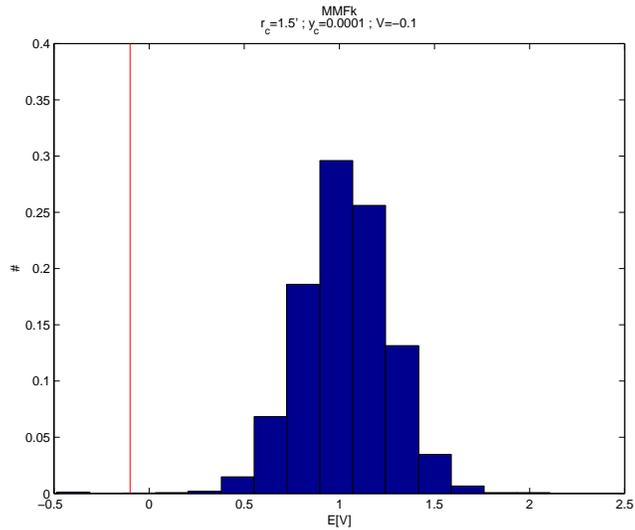}
\caption{Normalised histogram of the values of the estimated 
$V$ parameter using standard MMF for the kinematic SZ effect.
The true value of $V$ is shown with a vertical red line.}
\label{fig:fig3}
\end{figure}

Figure~\ref{fig:fig3} shows
the results of the estimation of $V$ for the 5000
simulated clusters using the standard MMF for the kinematic SZ effect.
The average value of the estimated $V$ is $\langle \hat{V} \rangle
= 1.05$ and the dispersion around this value is
$\sigma_{V} = 0.24$. The average standard deviation
of the filtered maps is, for comparison, $\sigma_{w_{MMFk}}/y_c=0.236$. 
The value for this quantity 
that was predicted in section~\ref{sec:preliminar} was $0.27$.
Eq. (\ref{eq:expected_biases}) predicts $\hat{V}/V = -9.5$, and the 
value that is found in the simulations is $\hat{V}/V = -10.5$.

The previous results have been obtained without previous 
knowledge of $y_c$, that is, the estimation of $V$ for each
simulated cluster is obtained by dividing the filtered map at
the position of the cluster by the value $\hat{y}_c$ estimated
with the MMFt. Therefore, the errors in the estimation
of $y_c$ do propagate and affect the estimation of $V$.
If we assume that $y_c$ is perfectly known we
can use is nominal value ($y_c=0.0001$ in this case)
and then we obtain $\langle \hat{V} \rangle = 1.03$ and 
$\sigma_{V} = 0.23$. We conclude that the uncertainty on
$y_c$ is relatively unimportant in this case.

\subsubsection{Performance of $\Phi$ multifilters}

\begin{figure}
\includegraphics[width=84mm]{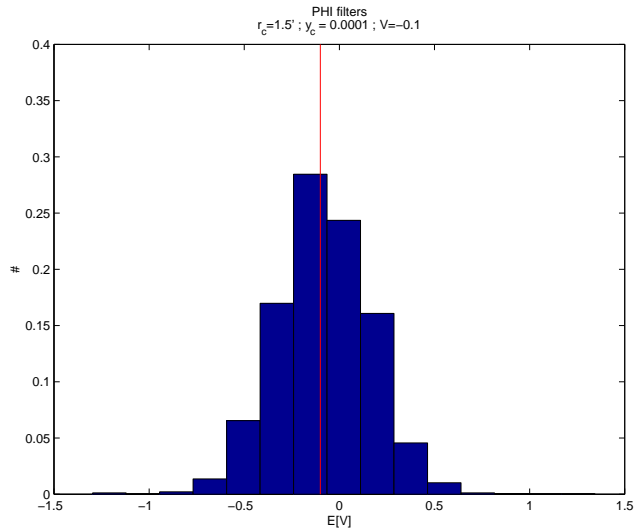}
\caption{Normalised histogram of the values of the estimated 
$V$ parameter using $\Phi$ multifilters for the kinematic SZ effect.
The true value of $V$ is shown with a vertical red line.}
\label{fig:fig4}
\end{figure}

Figure~\ref{fig:fig4} shows
the results of the estimation of $V$ for the 5000
simulated clusters using the $\Phi$ multifilters
for the kinematic SZ effect.
The average value of the estimated $V$ is $\langle \hat{V} \rangle
= -0.08$ and the dispersion around this value is
$\sigma_{V} = 0.26$. The average standard deviation
of the filtered maps is, for comparison, $\sigma_{w_{\Phi}}/y_c=0.249$. 
The value for this quantity 
that was predicted in section~\ref{sec:preliminar} was $0.28$.
The bias has been very much reduced, and the small deviation
from zero can in our opinion
be  attributed to the limited size of the sample and the 
large variance of the filtered maps.

If we assume that $y_c$ is perfectly known we
can use is nominal value ($y_c=0.0001$ in this case)
and then we obtain $\langle \hat{V} \rangle = -0.08$ and 
$\sigma_{V} = 0.25$. We conclude that the uncertainty on
$y_c$ is relatively unimportant in this case as well.

\subsection{Further tests on the $\Phi$ multifilters}

The toy clusters used above have large values of the comptonization $y_c$
and the peculiar velocity. Only a few clusters in the sky are expected
to have such large values of these parameters at the same time. Though
they serve well for our academic test, it is necessary to determine
if the filters are still unbiased in more realistic cases where the
cluster parameters take lower values. We performed two sets of
simulations varying the value of $y_c$ and $V$. The same as before,
for each value of $y_c$ and $V$ we simulated 5000 clusters, only
five of them per map. The core radius was fixed to $r_c=1$ pixel.

\begin{figure}
\includegraphics[width=84mm]{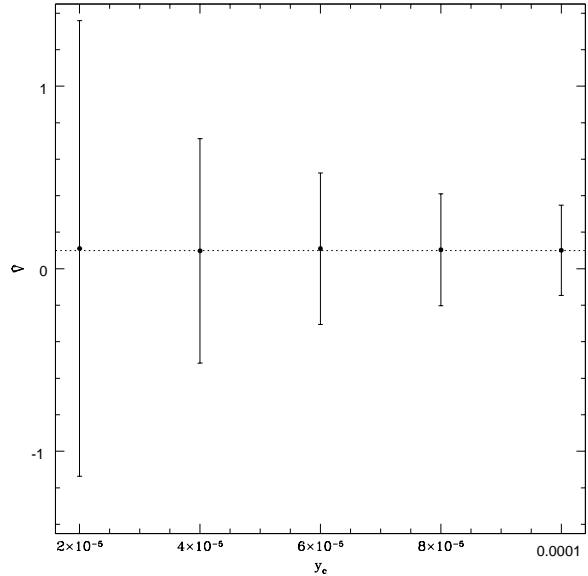}
\caption{Estimation of the $V$ parameter using $\Phi$ multifilters for different values 
of the comptonization $y_c$. The true value is indicated with an horizontal dotted line.}
\label{fig:fig5}
\end{figure}

Figure~\ref{fig:fig5} shows
how the $V$ parameter is estimated when varying the comptonization $y_c$. The true value of 
the velocity parameter was fixed to $V=0.1$. The estimation
is unbiased even for low values of $y_c$. The error bars grow as $y_c$ decreases, which
is not unexpected since $\sigma_V = \sigma_{w_{\Phi}} / y_c$.

\begin{figure}
\includegraphics[width=84mm]{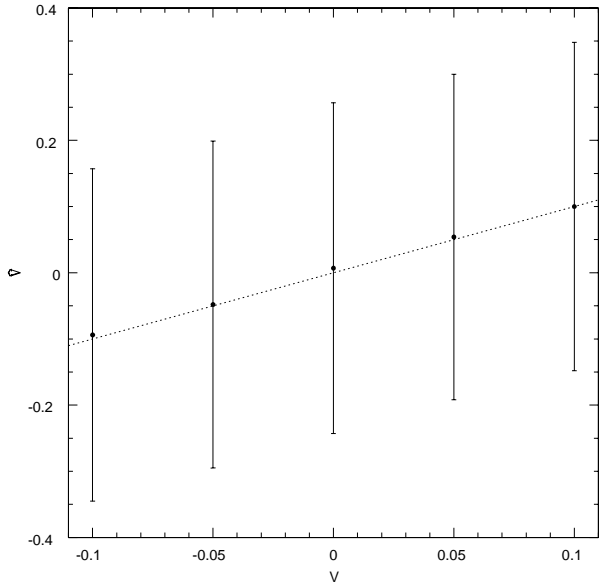}
\caption{Estimation of the $V$ parameter using $\Phi$ multifilters for different values 
of the true value of $V$. The true value is indicated with a dotted line.}
\label{fig:fig6}
\end{figure}

Figure~\ref{fig:fig6} shows
how the $V$ parameter is estimated when fixing $y_c=10^{-4}$ 
and changing the velocity of the cluster. The velocity can be
positive or negative.
The estimation
is unbiased for all the considered cases, even 
for the case $V=0$.

\section{Discussion} \label{sec:discussion}

In this work we have studied the performance of multifilters used to
enhance the Sunyaev-Zel'dovich effect signal in CMB maps such as the
future Planck satellite will obtain. The aim of multifilters is to
boost the cluster signal with respect to the background (CMB plus
other astrophysical sources plus noise) so that they can be more
easily detected and studied. 
Multifilters can be used
alone or  
as a previous step before more sophisticated data analysis tools
are applied to extract the maximum possible amount of information
from the data.

A problem arises when two different signals with different
frequency dependence but the same spatial profile appear in a 
given position. Such is the case of the thermal and kinematic SZ
effects. In that case, we have shown that standard matched
multifilters are intrinsically biased. This bias can be very
strong in the case of the weak kinematic SZ effect. We have
designed a family of unbiased matched multifilters that cancellate
this bias. The price paid for this bias cancellation is a small decrease
on the gain factor 
with respect to the standard matched multifilters. 
Unbiased matched multifilters use the \emph{a priori} knowledge on
both thermal and kinematic SZ effect's frequency dependences 
simultaneously in order to optimise the accuracy of the
estimation.
We have tested
the performance of both standard and unbiased matched multifilters
using realistic simulations of the nine Planck's frequencies. 
The simulations contain CMB, the main Galactic foregrounds, extragalactic
point sources and instrumental noise with the levels expected for
the different Planck channels. With these simulations we have shown
that standard matched multifilters can estimate the kinematic 
SZ effect with systematic biases much larger than the true value of
the effect. This problem is automatically solved using unbiased
matched multifilters.
The results of the numerical tests agree very well with
the theoretical expectations.

Our tests show that
the statistical error bars in the determination of the cluster
parameters $y_c$ and $V$ (or, equivalently, the cluster velocity $v_r$)
are directly related to the standard deviation of
the filtered map; in the case of the thermal SZ effect 
this last quantity does not depend on the intensity
of the SZ effect but on the properties of the background (i.e. the cross-power
spectrum matrix) and the shape of the filter (given by the source and beam profiles).
For the Planck sky patches here considered 
and using $\Psi$ multifilters, the typical error
bars in the determination of $y_c$ are $\sigma_{y_c} \sim 6 \times 10^{-6}$, independently
of the value of the comptonization of the clusters, which means 
that for ``bright'' clusters ($y_c \sim 10^{-4}$) individual comptonizations can
be determined with errors $\sim 20 \%$ using the Planck satellite and linear multifilters.
In the case of the kinematic SZ effect, to obtain $V$ ($v_r$) it is necessary
to divide the filtered map by $y_c$, so error bars are larger for 
low comptonizations (as reflected by figure~\ref{fig:fig5}).
For clusters as considered here, with $y_c=10^{-4}$ and temperature $T_e \simeq 5$ keV,
the $\Phi$ multifilters give statistical error bars $\sigma_V \simeq 0.26$, that is,
$\sigma_{v_r} \simeq 800$ \kms. The situation becomes worse for fainter clusters.
This means that Plank will not be able to tell us the velocities of individual 
clusters\footnote{At least if only multifilters are used to estimate the kinematic SZ effect.
Maybe the combination of the multifiltering technique --as a denoising step-- with more
sophisticated estimation methods already suggested in the literature could help to improve
these results. This is an exciting open possibility for future work.}.
Since $\Phi$ multifilters provide an unbiased estimator
of the kinematic SZ effect
it may be possible, however, to determine mean peculiar velocities on 
large scales by averaging over
many clusters. On the other hand, standard matched multifilters 
would lead to erroneous estimations of mean peculiar velocities due to
their intrinsic bias.

One of the strongest traits of linear filtering 
is that it requires only a small 
number of assumptions about the data to work. 
The multifilters
need only three information elements to do their work: 
the  frequency 
dependence of the SZ effects,
a reasonable knowledge on the generic cluster profile
(and the instrumental beam at each frequency)
and the cross-power spectra of the different channels'
background.
Uncertainties and errors in the \emph{a priori} information
must be kept to a minimum and, when present, they should be
taken into account.
In the following we will discuss
briefly some of the expected problematics that will arise in real
life clusters, and suggest a few ways
to deal with them.

\subsection{SZ frequency dependence}

The frequency dependence of the thermal and the kinematic SZ effects are 
very well known, so few surprises will come from this direction.  
Second-order relativistic corrections, however, can play a role, changing
the spectral shape of the effects. In particular, relativistic effects
can change the crossover frequency of the thermal SZ effect,
a region that is fundamental to the detection of the kinematic SZ effect.
The sensitivity of Planck will be too low to constrain the optical
depth of relativistic electrons $\tau_{rel}$ of 
individual clusters~\citep{en04}.
For other more sensitive experiments it may be necessary to take this effect
into account.
The relativistic correction to the thermal SZ effect frequency
dependence is given by~\citet{reph95} and it is a function of the 
frequency $\nu$ and the temperature of the electrons in the intracluster gas
$T_e$. 
The relativistic correction to the kinematic SZ effect 
is due to the Lorentz boost
to the electrons caused by the bulk velocity, introduces very small
spectral distortions on the kinematic SZ frequency dependence
and for a 10 keV cluster moving at 
1000 \kms the effect is about an 8 $\%$ of the non-relativistic term.
If $T_e$ is known, the relativistic corrections can be calculated,
eqs. (\ref{eq:SZt_freqdep}) and (\ref{eq:SZk_freqdep}) 
can be correspondingly modified and
the method can be used without problems. 

If $T_e$ is not known, it can be considered as another
quantity to be estimated. In principle, 
multifilters such as these here
described are able to estimate only amplitudes. However,
\citet{her02b} showed a way to estimate additional
parameters by means of consecutive
linear filterings. In that work, the additional parameter
to be estimated
was the core radius (as we will discuss later)
and the filter under study was the scale-adaptive filter.
It is straightforward to generalise the idea to UMMF and
other parameters such as the frequency dependence. The 
philosophy of the method is as follows: it can be shown
that matched filters (an unbiased matched filters) give a
maximum amplification of the sources when the correlation
between the shape of the filter and the shape of the
source is perfect. If for example the shape (profile) of
the source is known except for a scale factor (such as
the core radius), we can filter the image with a number
of different matched filters with varying core radius parameter,
and study how amplification changes with $r_c$. The value
of the test parameter at which the maximum amplification is
obtained will be our estimate of $r_c$. The same applies
to multifilters: the maximum amplification will be obtained
when the frequency dependence assumed for the filter matches
the true one. Then, we can pass a set of different multifilters
with varying frequency dependences (in this case, parametrised
by different values of $T_e$). The maximum amplification occurs when
the value of $T_e$ 
of the multifilter coincides with the true one. 

\subsection{Cluster profiles}

For the sake of clarity we have deliberately kept the examples presented here
as very simple, academic tests,
at least  
regarding to our SZ cluster simulations.
On the other hand, the background simulations --CMB,
Galactic foregrounds, extragalactic point sources 
and Planck instrumental noise levels as well as
pixel and beam sizes-- are very realistic. The purpose of
this is to focus on the effect of the ``contaminants'' rather
than the whole problem. 
Unfortunately, in real life 
clusters are not so nicely symmetric and
we do not have a perfect \emph{a priori}
knowledge of their spatial profile.
Luckily, we neither have a total ignorance about it.
Normally we have some previous idea about how clusters look like.
The main uncertainties here are:

\subsubsection{Size uncertainties}

Clusters of many different sizes do appear in CMB maps. Throughout the 
examples presented in this work we assumed that all the clusters had
the same core radius $r_c$, which was known a priori.~\citet{her02b} 
provided a method to deal with this uncertainty using the scale-adaptive
filter on a single map. The generalization to multifilters is
straightforward \citep{her02}.
The basic idea is to filter the maps
with a set of
different filters varying the $r_c$
parameter and then to look for the maximum gain cluster by cluster. 
With this method it is possible
to determine the value of $r_c$ with typical
errors lower than the pixel
size~\citep{her02b}. 

\subsubsection{Profile shape}

If instead of the modified multiquadric profile used in eq. 
(\ref{eq:cluster_profile}) the cluster has a different profile
the matching between the filter and the cluster will not be 
perfect and the filtering will loose performance. This will
be an important issue in high resolution experiments where the
cluster size is much larger than the beam size. In the case of
Planck most clusters will be smaller than the beam size
and after convolution their
observed shapes will be 
dominated by the beam profile. Therefore, in the case of
Planck the errors due to profile uncertainties will be
small except for a few cases of very extense clusters.

For these few extense cluster and for the case of
high resolution experiments we suggest the use of an iterative 
technique to adaptively fit the 
true profile of a cluster in an empiric way. Starting with a
given profile guess such as the one in eq. (\ref{eq:cluster_profile}),
the shape of the profile can be slowly varied in successive 
steps, introducing small filter shape perturbations
and maximising the signal to noise gain 
after filtering. The maximum gain corresponds to the best fit
between the profile of the filter and the profile of the cluster.
A method for adaptive shape fitting that could be adapted to
this problem has been
successfully accomplished in the modelling of 
gravitational 
lensing effect due to massive galaxy clusters (see for 
example~\citet{tom}). We will explore this possibility in
a future work.

\subsubsection{Non-circular profiles} 

In this work
we have assumed symmetric circular profiles and beams. 
This condition
can be relaxed without loss of generality. If the non-symmetric profile is
previously known (for example thanks to high-resolution observations
in X-ray or other wavelengths) specific non-symmetric filters can be
calculated without problem. If the shape of the cluster is not previously 
known, an iterative scheme such as the previously mentioned can
be attempted.

\subsubsection{Non iso-thermal clusters}

It is frequent to consider that $T_e$ is constant all along the cluster,
which allows us to write nicely the kinematic SZ effect as in eq. (\ref{eq:SZk_freqdep}).
But galaxy clusters are not iso-thermal. As a consequence of this, cluster parameters
derived through the observation of SZ effect will be affected.~\citet{han04b} has shown
that peculiar velocities will be systematically shifted by $10 - 20 \%$ if the 
inner structure of the cluster is not considered in detail, which requires either 
observationally expensive X-ray observations or high angular resolution SZ observations.  

\subsection{Background power spectrum}

The background power spectrum 
can usually be
estimated directly from the data. 
This
is a safe practise if
SZ clusters' contribution to the total power spectrum
is relatively small. Such is the case expected for the Planck mission.
If the
SZ clusters were so strong that they contributed significantly to the
total power spectrum at all the scales, they would be so conspicuous
that detection would not be a problem, and multifiltering would not 
be necessary.

\subsection{Extragalactic point source contamination}

Historically, extragalactic point sources have been the 
strongest source of contamination in SZ observations.  
In this work we have included realistic point source 
simulations using the~\citet{T98} model,
whose predictions for the radio source counts have
been recently confirmed by the WMAP mission~\citep{bennett}.
We have shown that multifilter estimation is not
substantially affected by these point sources. There are however
two details that have not been included in the simulations and
can affect the estimation of cluster parameters: the intrinsic
galaxy clustering and the spatial correlation between individual
galaxies and galaxy clusters.

\subsubsection{Point source clustering}

We used the point sources of the~\citet{T98} model assuming a uniform Poisson
distribution of galaxies in the sky. In the real sky, however, galaxies are 
spatially correlated due to clustering. 
Temperature fluctuations due to unresolved point sources are stronger
if clustering is considered. Apart from increasing the error bars, 
~\citet{nabila04a}
have shown that this effect leads to 
significant systematic errors in the determination of the 
thermal and kinematic SZ amplitudes by means
of the SASZ method~\citep{han04a}. It is not clear how galaxy clustering
will affect the multifiltering method presented in this work. 
If the correlation scale of the galaxies is similar to the
cluster size systematic shifts can occur\footnote{This is due
to the same kind of effect 
that is the topic of this work: a cluster-like structure (in this case
a ``lump'' of correlated galaxies with a correlation scale approximately
equal to the cluster size) 
is superimposed
to the observed cluster, showing a different frequency dependence but
nevertheless producing a bias in the estimation of the cluster parameters.},
but if the unresolved galaxies' fluctuation field is not correlated with the 
positions of the clusters the effect should be small. This is a problem worth
to study in future works.

\subsubsection{Point sources associated to clusters}

Another factor not considered in the simulations is the associations 
between clusters and individual galaxies. These associations can be
due to the galaxies that form part of the cluster or to gravitational
lensing, that increases the number of faint galaxies that are observed
in the direction of the cluster through magnification effects.
Emission from such galaxies can fill in the SZE decrement, leading
to a wrong estimation of the effect.
Galaxies
inside the cluster are likely to affect the estimation of SZ effect
at high frequencies while gravitational lensing effect is expected
to be an issue for frequencies $< 30$ GHz~\citep{car02}. The solution
to this problem is increasing the resolution of the observations
so that individual galaxies can be detected an accounted for; in
low-resolution experiments such as Planck this will not be possible.
The different frequency dependence of the SZ effect and galaxy
emission will help to reduce contamination, but some residual
effects will remain that can lead to additional unavoidable biases. 

We remark that the method presented in this work is not
incompatibly with other SZ detection and estimation
techniques. Linear filtering is an useful tool that can be used
alone or as a step inside a more ambitious analysis of the data.
The filters here introduced are computationally fast, robust,
efficient and unbiased. They reduce the contamination due to 
noise and other astrophysical emission and optimise the separation
between the thermal and the kinematic effect using all the frequency 
channels available (and not only the channels around the thermal
crossover frequency). In the context of the future Planck mission, 
the implementation of the filters is straightforward, but due to the
limitations on angular resolution and sensitivity the kinematic SZ
effect will be badly determined. Experiments with more sensitivity
and angular resolution will be much better for the detection of
this elusive effect, but the implementation of the filters 
will require much more care since the inner structure of the clusters
will reveal its complexity. We have hinted in the discussion some
ways to deal with this complexity; applications to specific cases
will be addressed in future works.

\section*{Acknowledgments}

The authors thank Patricio Vielva for useful suggestions.
DH acknowledges support from 
the
European Community's Human Potential Programme
under contract HPRN-CT-2000-00124, CMBNET, and the 
Instituto de F\'\i sica de Cantabria for the hospitality
during several research stays in 2003. 
RBB thanks the Spanish Ministerio de Ciencia y Tecnolog\'\i a and
the Universidad de Cantabria for a Ram\'on y Cajal contract.
MC acknowledges support from the Spanish Ministerio de Educaci\'on
for a doctoral FPI fellowship.
We thank FEDER Project 1FD97-1769-C04-01, Spanish DGESIC Project 
PB98-0531-C02-01 and INTAS Project INTAS-OPEN-97-1192 for 
partial financial support.

\label{lastpage}

\end{document}